\documentclass[a4paper,10pt]{article}
\usepackage[dvips]{graphicx}
\usepackage{amssymb,amsmath}
\numberwithin{equation}{section}

\begin{document}
\title{\bf FRW cosmology of the generalized model of  LQG}
\author{{S. Chattopadhyay$^{a}$\thanks{Email: surajcha@iucaa.ernet.in; surajitchatto@outlook.com} , \hspace{1mm} A. Ashurov$^b$, \hspace{1mm} M. Khurshudyan$^{c}$\thanks{Email:
khurshudyan@yandex.ru} ,  \hspace{1mm} K. Myrzakulov$^b$ } \\\\
and\\\\  {A. Pasqua$^{d}$\thanks{Email: toto.pasqua@gmail.com} , \hspace{1mm} R. Myrzakulov$^{b}$\thanks{Email: rmyrzakulov@gmail.com; rmyrzakulov@csufresno.edu}}\\\\
$^{a}${\small {\em Pailan College of Management and Technology, Bengal Pailan Park, Kolkata-700 104, India}}\\
$^{b}${\small {\em Eurasian International Center for Theoretical Physics and  Department of General }} \\ 
{\small {\em $\&$  Theoretical Physics, Eurasian National University, Astana 010008, Kazakhstan }}\\
$^{c}${\small {\em Department of Theoretical Physics, Yerevan State
University, 1 Alex Manookian, Yerevan, Armenia}}\\
$^{d}${\small {\em Department of Physics, University of Trieste, Via Valerio, 2 34127 Trieste, Italy}}\\ } 

%\date{}

\date{}
\maketitle
\begin{abstract}
In this paper, we study the main cosmological properties of the classical Friedmann equations in the case of homogeneous and isotropic Friedmann-Robertson-Walker Universe and we also generalized the expression of the Friedmann equation in the case of Loop Quantum Cosmology (LQC). Considering the $M_{35}$-model, we found the solutions of the equations considered for two particular cases, i.e. $Q=0$ (i.e., the de Sitter solution) and $Q>0$. Moreover, we considered and studied two exact cosmological solutions of the $M_{35}$-model, in particular the power-law and the exponential ones. Futhermore, we also considered a third more complicated case and we derived the solution for an arbitrary function of the time $f\left(t\right)$. A scalar field description of the model is presented  by constructing its self-interacting potential.
\end{abstract}
\vspace{2cm}

\sloppy

\tableofcontents

\section{Introduction}

One of the long-standing problems in the standard Big Bang cosmology is the initial singularity from which all matter and energy originated.
Standard cosmology based on General Relativity offers no resolution to this problem. However, a quantum gravitational model of Loop Quantum Gravity
(LQG) offers a nice solution. The theory and principles of LQG, when applied in the cosmological framework, creates a new theoretical framework of Loop
Quantum Cosmology (LQC) \cite{1}.

The main idea is that LQC assumes a discrete nature of space which leads, at quantum level, to consider a Hilbert space where quantum states are
represented by almost periodic functions of the dynamical part of the connection \cite{11}.

Due to quantum corrections, the Friedmann equations get modified. The Big Bang singularity is resolved and replaced by a quantum bounce \cite{6}.
For a brief summary on Loop Quantum Cosmology, see Reference \cite{7}.

Nowadays, the accelerated expansion of the present universe has been supported by various independent cosmological observations. As representative
procedures to explain the late time acceleration, the first is to assume the existence of an unknown component called Dark Energy,
which has as one of the main feature a negative pressure (which can be considered as responsible for the accelerated expansions the Universe is
undergoing). The second is to modify gravity, the simplest model of which is $f(R)$ gravity.

In general, Dark Energy can be assumed to be a perfect fluid with equation of state given by $P=\rho-f(\rho)$, which realizes the current cosmic
acceleration. Moreover, the Wilkinson Microwave Anisotropy Probe (WMAP) observations indicate that the central value of the equation of state (EoS)
parameter $\omega$ is given by $\omega\equiv P/\rho \approx - 1.10$ \cite{10}. This means that our universe would be dominated by phantom energy
 ($f(\rho)\geq 0$, i.e.  $\omega < -1$).

The standard viewpoint of LQC assumes, at quantum level, a discrete nature of space which leads to a quadratic modification $(\rho^2)$ in its effective
Friedmann equation at high energies \cite{4}. This modified Friedmann equation depicts the ellipse in the plane $(H; \rho)$, where $H$ is the Hubble
parameter and $\rho$ the energy density of the universe (for more details, see Reference \cite{5}).

We must also underline here that the LQC model prevents singularities like the Big Bang or the Big Rip. Using LQC when one considers a model of universe filled by radiation and matter where, due to the cosmological constant, there are a de Sitter and an anti de Sitter solution \cite{2}.
	
The effects of Loop Quantum Gravity can be described in two possible ways: the first one is based on the modification of the behavior of the inverse scale factor operator. This approach has been used to study quantum bounces, avoidance of singularities and to produce inflationary expansion \cite{8}. A second approach is to add a term quadratic in density to the Friedmann equation. In LQC, the non-perturbative effects lead to correction term $-\rho^2/\rho_c$ to the standard Friedmann equation. With the inclusion of this term, the Universe bounces quantum mechanically as the energy density of matter-energy reaches the level of $c$ (i.e. of order of the Planck density). Thus the LQC is non-singular by producing a bounce before the occurrence of any potential singularity and hence transitions from a pre-Bang and after-Bang are all well-defined. The observational constraints due to the quadratic term are discussed in scientific literature [9] where it is shown that the model with quadratic correction to density is consistent with the observational tests.

The development of LQC as a $f(T)$ theory allows the study of LQC perturbations using the perturbation equations in $f(T)$ modified gravity. This is an alternative to the study of perturbations in LQC up to the present, which is based on phenomenological corrections.

In Reference \cite{3}, LQC have been considered and the modified Friedmann and Raychauduri equations has been derived. Moreover, future singularities in LQC have been studied and it was shown that the Rip singularities do not survive, but Type II and Type IV singularities could still happen.

For other models in ordinary $f(T)$ gravity, namely, not in the framework of LQC, the Type I and Type IV singularities can eventually appear in the finite time limit for a power-law form of $f(T)$. Moreover, the LR and PL scenarios can be realized for specific power-law type models of $f(T)$ gravity \cite{12}. Accordingly, the features of future singularities occurring in $f(T)$ gravity in the context of LQC would be different from those of other models in $f(T)$ gravity.

In this paper, we consider modified teleparallel gravity theory with the torsion scalar have recently gained a lot of attention as a possible explanation of Dark Energy. We perform a thorough reconstruction analysis on the $f(T)$ models, where $f(T)$ is some general function of the torsion term, and derive conditions for the equivalence between of $f(T)$ models with purely kinetic k-essence. We present a new class models of $f(T)$-gravity and k-essence. We also proposed some new models of generalized gases and knot universes as well as some generalizations of $f(T)$ gravity.

In this paper, we will investigate the evolution
of our universe dominated by a scalar field in LQC, which has constant equation of state
and interacts with Dark Matter, and then investigate whether there are some interesting
features arising from the loop quantum gravity effect. We  will concentrate to the Fridmann-Robertson-Walker metric case of the form:
\begin{eqnarray} \label{frw}
ds^2=dt^2-a(t)^{2}(dx^2+dy^2+dz^2),
\end{eqnarray}
where $a(t)$  is the scale factor.

This paper is organized in the following way.
In Section 2, we give a brief description of the classical Friedmann-Robertson-Walker (FRW) model.
In Section 3, we present a brief review of  the standard FRW model of Loop Quantum Gravity (LQG).
In Section 4, we study two particular cases of the $M_{35}$-model.
In Section 5, we study two exact cosmological solution of the $M_{35}$-model, in particular the power-law and the exponential ones.
In Section 6, we study the scalar field analog of the $M_{35}$-model.
Finally, in Section 7 we write the Conclusions of this paper.

\section{Brief review of the classical FRW model}

In this Section, we would like first to give a brief review  of the classical FRW model of General Relativity. These results are well-known and so that we present them without citation (see, review paper on cosmology). Let us we consider the  Einstein-Hilbert action of the form
\begin{eqnarray} \label{eh}
S=\int \sqrt{-g}d^4x(R+L_m),
\end{eqnarray}
where $R$ is the curvature scalar, $L_m$ is the matter Lagrangian and $g=\det{g_{ij}}$ is the determinant of the metric. Then, the classical Friedmann equations for homogeneous and isotropic Friedmann-Robertson-Walker (FRW) models of the Universe are given by:
\begin{eqnarray} \label{fe1}
3H^2&=&8\pi G\rho,\\
\label{fe2} \dot{H}&=&-4\pi G(\rho+p),\\
\label{fe3} \dot{\rho}&=&-3H(\rho+p).
\end{eqnarray}
From Eqs. (\ref{fe1}) and (\ref{fe3}), we get the energy density $\rho$ and its first time derivative $\dot{\rho}$ as follow:
\begin{eqnarray}
\label{rat1} \rho&=&\frac{3H^2}{8 \pi G}, \\
\label{rat2} \dot{\rho}&=&\frac{3H \dot{H}}{4\pi G}.
\end{eqnarray}
Moreover, from Eq. (\ref{rat3}), we can define the following expression for the pressure $p$:
\begin{equation}
p=-\left(\rho+\frac{\dot{\rho}}{3H}\right). \label{rat3}
\end{equation}
Using Eq. (\ref{rat3}), we have that the equation of state parameter $\omega$ can be written as:
\begin{equation}
\omega=\frac{p}{\rho}=-1-\frac{\dot{\rho}}{3H\rho}.
\end{equation}
For a pedagogical reasons and for the self-contained aim with the next Sections, we now consider the well-known solution (obtained by other Authors) corresponding to:
\begin{eqnarray}
\ln a=\alpha t^n+\beta. \label{rat4}
\end{eqnarray}
\begin{figure}[h!]
\begin{center}
\begin{minipage}[h]{0.4\linewidth}
\includegraphics[width=1\linewidth]{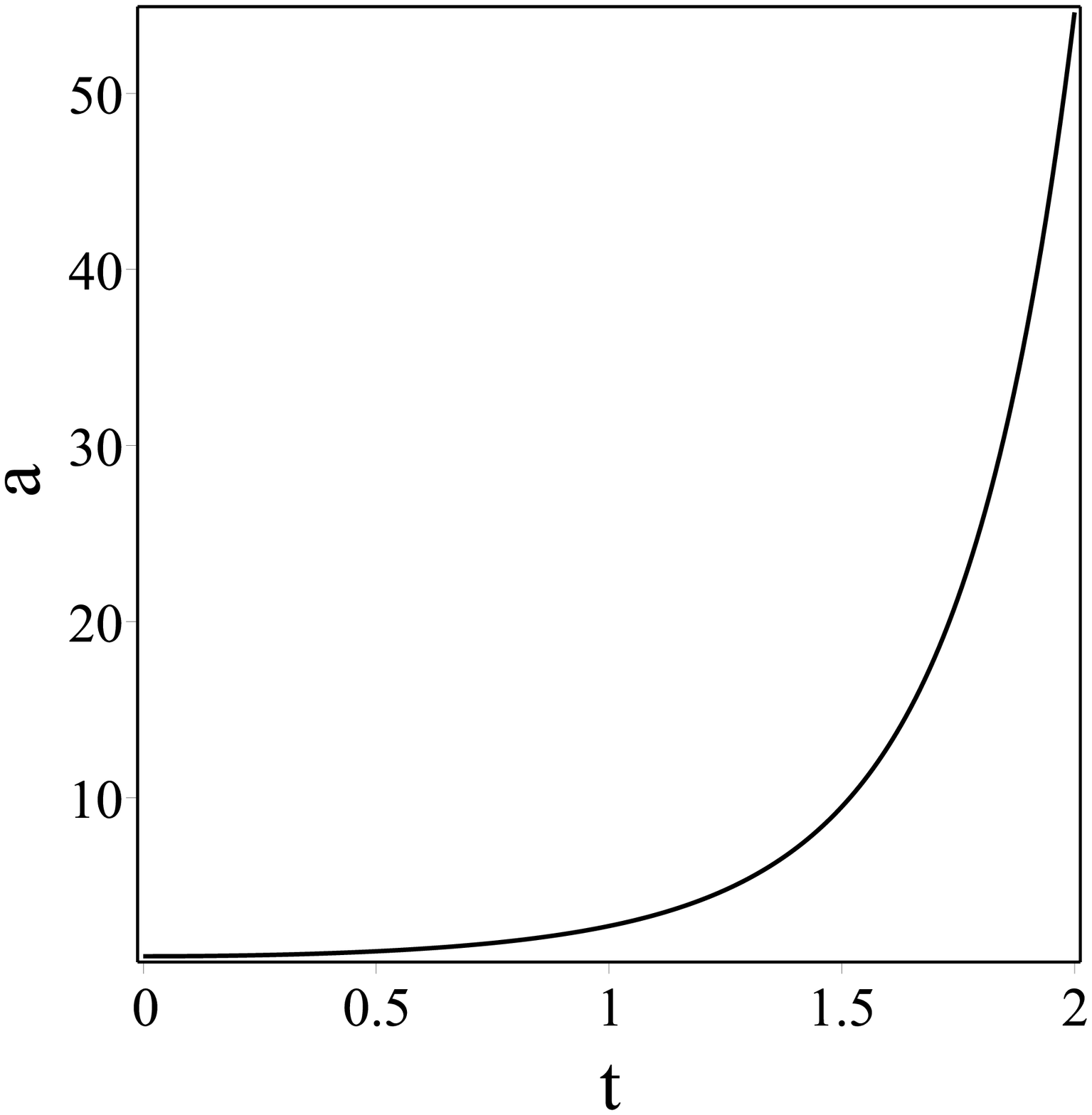}
\caption{Plot of the scale factor $a$ given in Eq. (\ref{rat5}) as a function of the cosmic time $t$.} %% ������� � �������
\label{kz1} %% ����� ������� ��� ������ �� ����
\end{minipage}
\hfill
\begin{minipage}[h]{0.4\linewidth}
\includegraphics[width=1\linewidth]{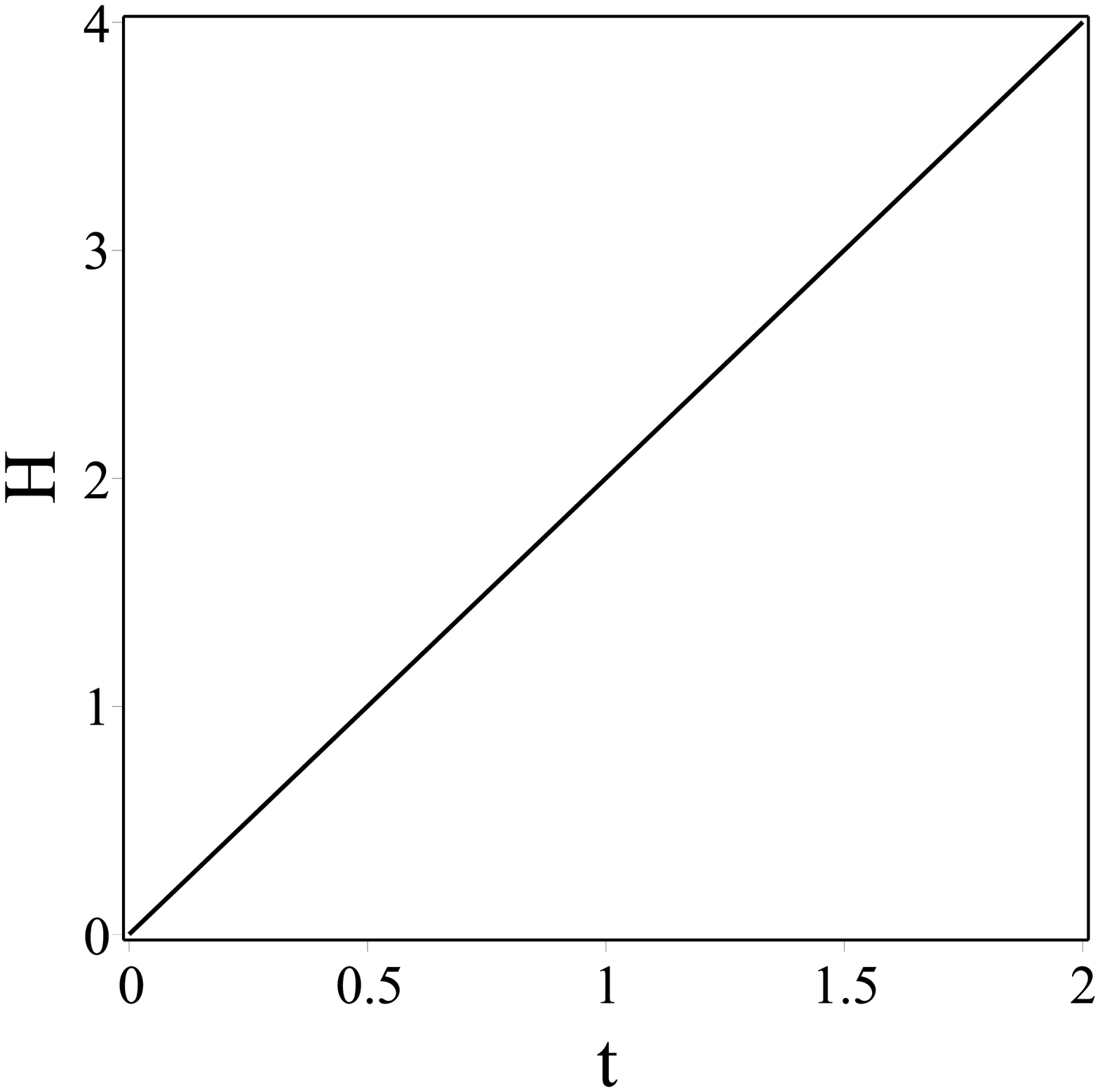}
\caption{Plot of the Habble parameter $H$ given in Eq. (\ref{rat6}) as a function of the cosmic time $t$.}
\label{kz2}
\end{minipage}
\end{center}
\end{figure}
Then, we obtain the following expressions for the scale factor $a$, the Hubble parameter $H$ and the first time derivative of the Hubble parameter:
\begin{eqnarray}
a&=&a_0 e^{\alpha t^n}, \label{rat5} \\
H&=&\frac{\dot{a}}{a}=n\alpha t^{n-1},\label{rat6} \\
\dot{H}&=&n(n-1)\alpha t^{n-2},\label{rat7}
\end{eqnarray}
where $a_0=e^{\beta}$.

\begin{figure}[h!]
\begin{center}
\begin{minipage}[h]{0.4\linewidth}
\includegraphics[width=1\linewidth]{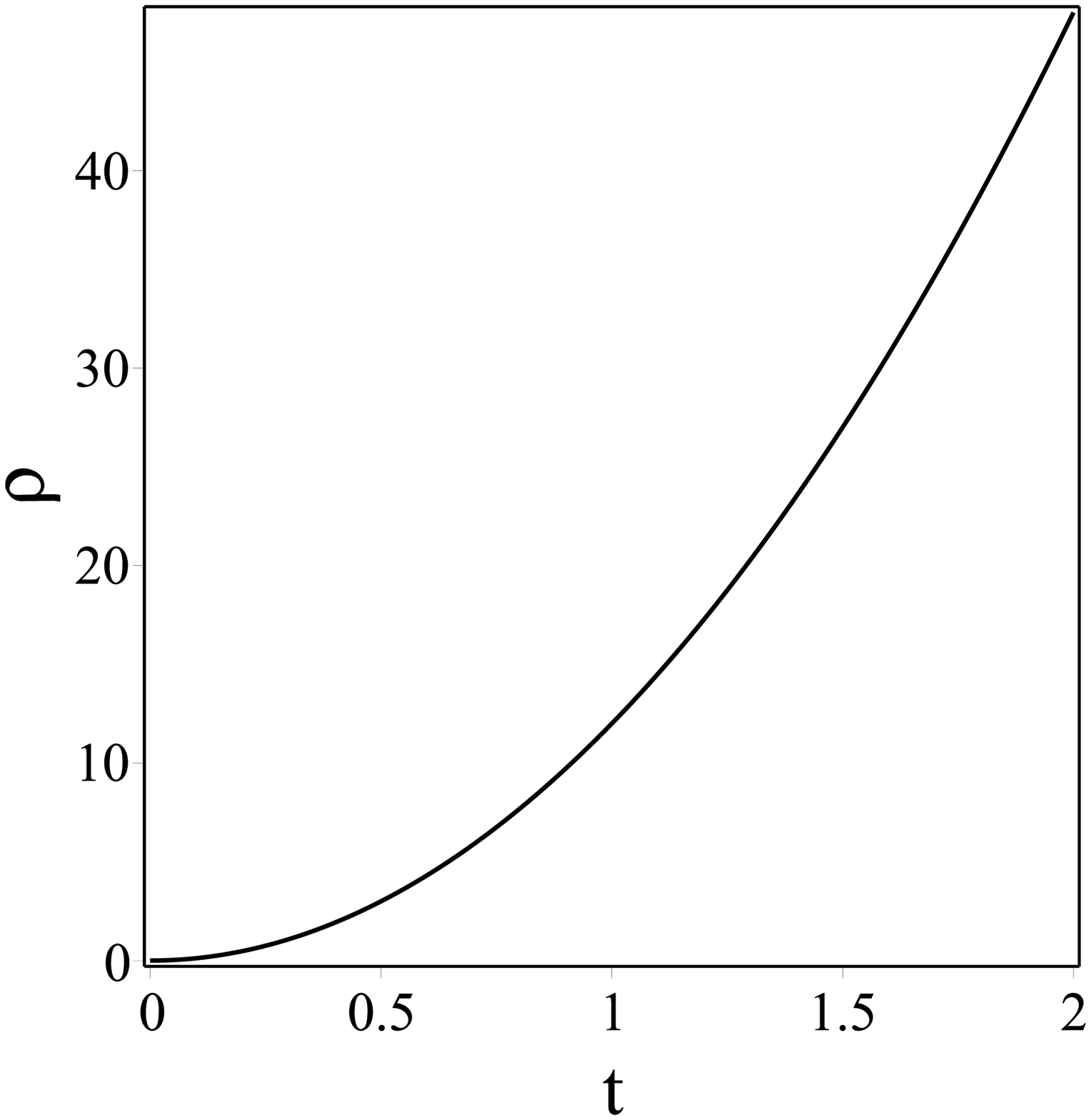}
\caption{Plot of the energy density $\rho$ given in Eq. (\ref{rat3}) as a function of the cosmic time $t$.} %% ������� � �������
\label{kz3} %% ����� ������� ��� ������ �� ����
\end{minipage}
\hfill
\begin{minipage}[h]{0.4\linewidth}
\includegraphics[width=1\linewidth]{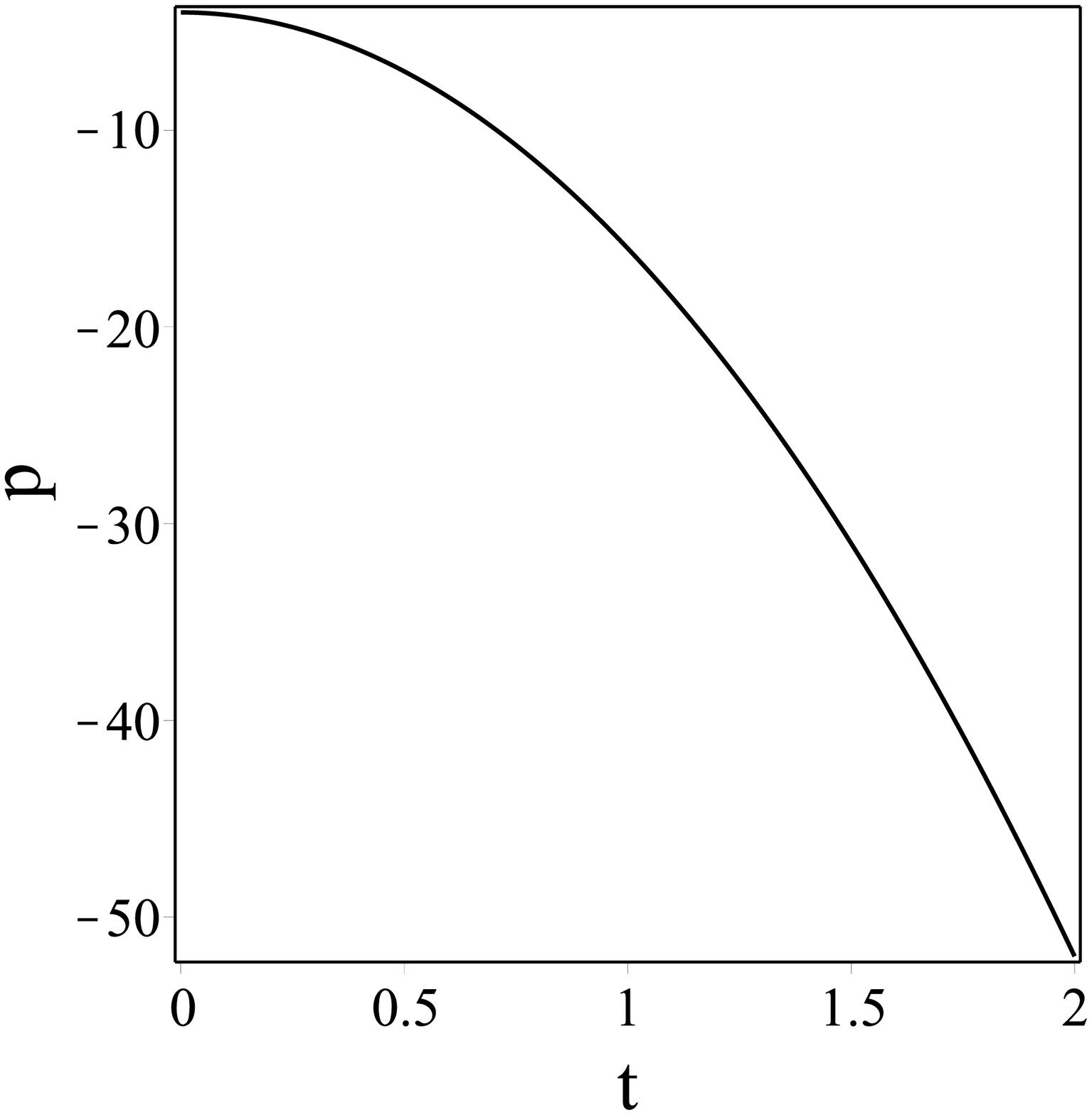}
\caption{Plot of the pressure $p$ given in Eq. (\ref{rat6!}) as a function of the cosmic time $t$.}
\label{kz4}
\end{minipage}
\end{center}
\end{figure}
\begin{figure}[h!]
	\centering
		\includegraphics[width=7cm]{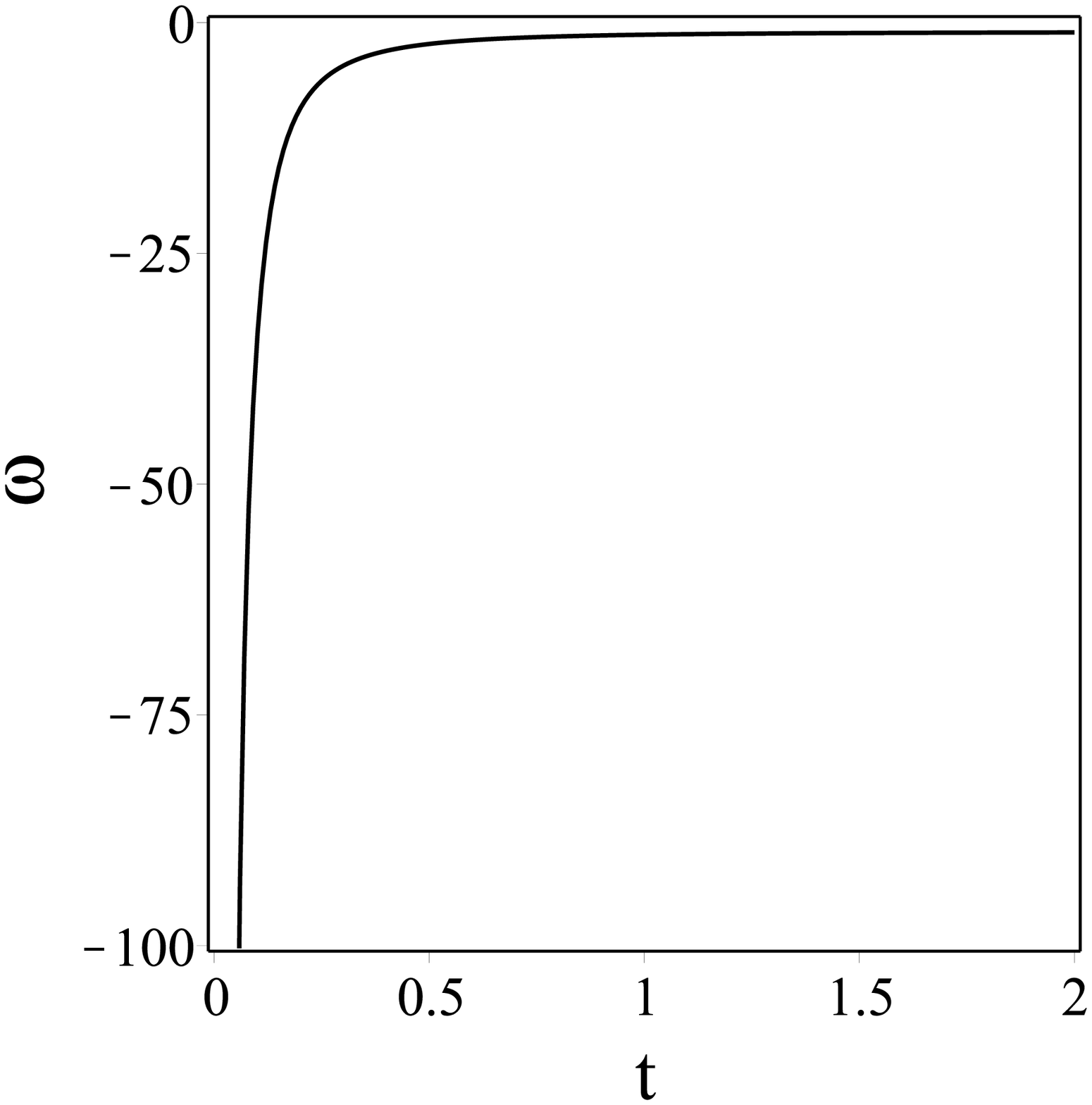}
		\caption{Plot of the EoS parameter $\omega$ given in Eq. (\ref{rat6!!!}) as a function of the cosmic time $t$.}
	\label{kz5}
\end{figure}
Moreover, using the expressions given in Eqs. (\ref{rat5}), (\ref{rat6}) and (\ref{rat7}) in Eqs. (\ref{rat1}), (\ref{rat3}) and (\ref{rat4}), we obtain the following quantities:
\begin{eqnarray}
\rho &=& \frac{3n^2 \alpha ^2}{8\pi G} t^{2(n-1)},\label{rat6!} \\
\dot{\rho} &=& \frac{3n^2 \alpha ^2 (n-1)}{4 \pi G} t^{2n-3}, \label{rat6!!}\\
\omega &=& -\left[1+\frac{2(n-1)}{3n\alpha}\cdot t^{-n}\right].\label{rat6!!!}
\end{eqnarray}
Analyzing the quantities obtained above, we can conclude that $\omega\rightarrow -1$ for $t\rightarrow \infty$, which means that the Universe is on the accelerated expansion phase.

In Fig. \ref{kz1} and Fig. \ref{kz2}, we show the cosmological evolution of the scale factor $a(t)$ and the Hubble parameter $H$ as a function of the cosmic time $t$. We also depict the cosmological evolutions of the energy density $\rho$ and the pressure $p$ as functions of the cosmic time $t$ in Fig. \ref{kz3} and Fig. \ref{kz4}. Furthermore, in Fig. \ref{kz5}, we demonstrate the cosmological evolution of the EoS parameter $\omega$ as a function of the cosmic time $t$. The parameters for the model considered are chosen as $\alpha=1$, $n=2$ and $a_0 = 1$.

%%%%%%%%%%%%%%%%%%%%%%%%%%%%%%%%%%%%%%%%%%%%%%%%%%%%%
\section{Brief review of the standard FRW model of LQG}
%%%%%%%%%%%%%%%%%%%%%%%%%%%%%%%%%%%%%%%%%%%%%%%%%
In this Section, we want to focus our attention to the main features of the standard FRW model of LQG. We present these  well-known  results (see i.e.,  \cite{LQG1}-\cite{LQG2}) here to be self-contained and to fix our notations. In the LQG, the classical Friedmann equations:
\begin{eqnarray}
\left(\frac{\dot{a}}{a}\right)^{2}&=&\frac{8\pi G}{3}\rho-\frac{k}{a^2}+\frac{\Lambda}{3}, \label{fe1a} \\
 \frac{\ddot{a}}{a}&=&-\frac{4\pi G}{3}(\rho+3p), \label{fe1b}
\end{eqnarray}
get corrected by the factor:
 \begin{eqnarray}
\left(\frac{\dot{a}}{a}\right)^{2}&=&\frac{8\pi G}{3}\rho\left(1-\frac{\rho}{\rho_{c}}\right)-\frac{k}{a^2}+\frac{\Lambda}{3}, \label{fe1c} \\
 \frac{\ddot{a}}{a}&=&-\frac{4\pi G}{3}\left[\rho\left(1-\frac{4\rho}{\rho_{c}}\right)+3p\left(1-\frac{2\rho}{\rho_{c}}\right)+\frac{\Lambda}{3}\right], \label{fe1d}
\end{eqnarray}
where the critical energy density $\rho_c$ is given by:
\begin{eqnarray}
\rho_{c}=\frac{8\pi G}{3}\left(\gamma^{2}a_{0}\right)^{-1}. \label{fe1e}
\end{eqnarray}
Then, in terms of the Hubble parameter $H$, these  modified Friedmann equations of the  LQC take the following forms (we here assumed $k=\Lambda=0$) \cite{LQG3}-\cite{LQG4}:
\begin{eqnarray} \label{fe1}
3H^2&=&8\pi G\rho\left(1-\frac{\rho}{\rho_c}\right),\\
 \label{fe2} \dot{H}&=&-4\pi G(\rho+p)\left(1-\frac{2\rho}{\rho_c}\right),\\
 \label{fe3}\dot{\rho}&=&-3H(\rho+p),
\end{eqnarray}
where we used the units $8\pi G = 1$ (where $G$ represents the Newton's gravitational constant), $\rho$ is the total cosmic energy density, $\rho_{c} = \frac{\sqrt{3}}{16 \pi^{2} \gamma^{3} G^{2} \hbar}$ denotes the critical Loop Quantum density and $\gamma$ is the dimensionless Barbero-Immirzin parameter (it is suggested that $\gamma = 0.2375$ by the black hole thermodynamics in LQG \cite{13}).
From Eq. (\ref{fe1}), we can obtain the following expression:
\begin{eqnarray}
\rho^2-\rho \rho_c+k_F H^2=0,  \label{fe4}
\end{eqnarray}
where $k_F=\frac{3\rho_c}{8\pi G}$.
The solutions of Eq. (3.4) are given by:
\begin{eqnarray}
\rho_1 &=& \frac{\rho_c+\sqrt{D}}{2}, \label{fe5}\\
\rho_2 &=& \frac{\rho_c-\sqrt{D}}{2},  \label{fe6}
\end{eqnarray}
where:
\begin{eqnarray}
D=\rho_c^2-4k_F H^2 \geq 0. \label{fe7}
\end{eqnarray}
Then, in order to have real solution for Eqs. (\ref{fe5}) and (\ref{fe6}), the Hubble parameter squared $H^2$ must satisfy the following condition:
\begin{eqnarray}
H^2 \leq \frac{\rho_c^2}{4k_F}. \label{fe8}
\end{eqnarray}
Taking into account Eqs. (\ref{fe5})-(\ref{fe7}), we can write
\begin{eqnarray}
\dot{\rho_1} &=&-2k_F H \dot{H} D^{-1/2}, \label{fe9}\\
\dot{\rho_2} &=& 2k_F H \dot{H} D^{-1/2}. \label{fe10}
\end{eqnarray}

\begin{figure}[h]
\begin{center}
\begin{minipage}[h]{0.4\linewidth}
\includegraphics[width=1\linewidth]{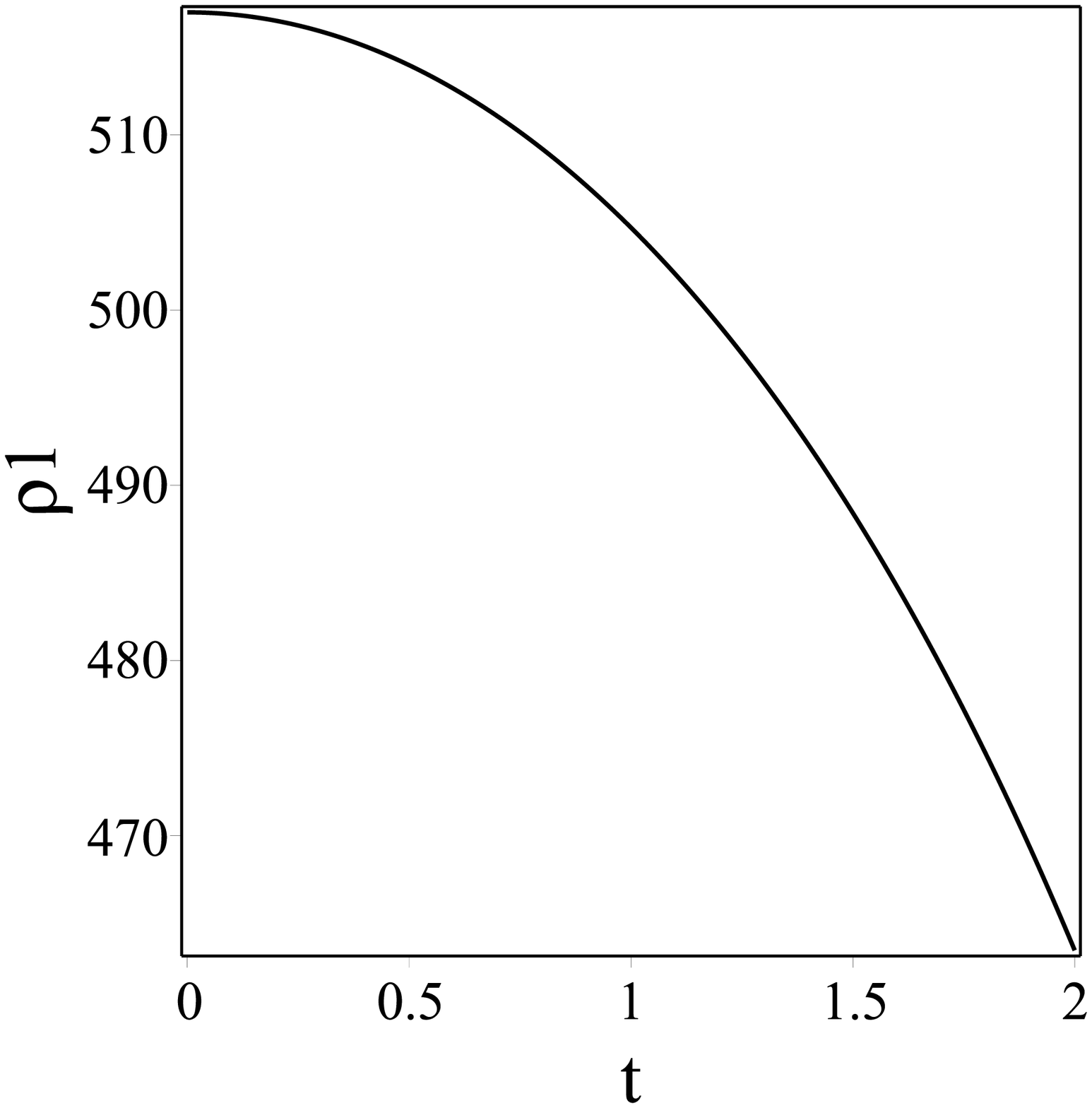}
\caption{Plot of the energy density $\rho_1$ given in Eq. (\ref{fe5}) as a function of the cosmic time $t$.} %% ������� � �������
\label{kz6} %% ����� ������� ��� ������ �� ����
\end{minipage}
\hfill
\begin{minipage}[h]{0.4\linewidth}
\includegraphics[width=1\linewidth]{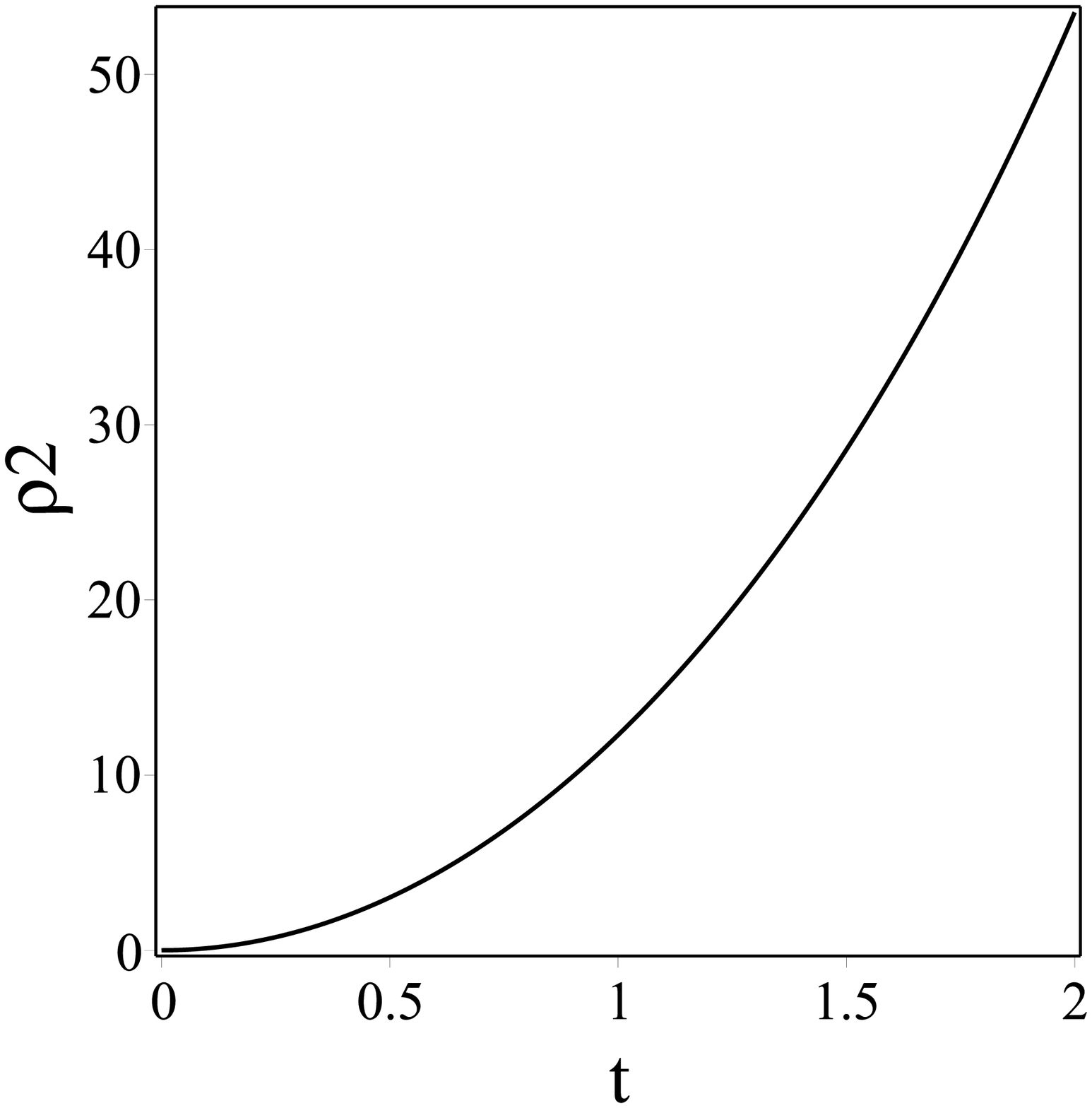}
\caption{Plot of the energy density $\rho_2$ given in Eq. (\ref{fe6}) as a function of the cosmic time $t$.}
\label{kz7}
\end{minipage}
\end{center}
\end{figure}

\begin{figure}[h]
\begin{center}
\begin{minipage}[h]{0.4\linewidth}
\includegraphics[width=1\linewidth]{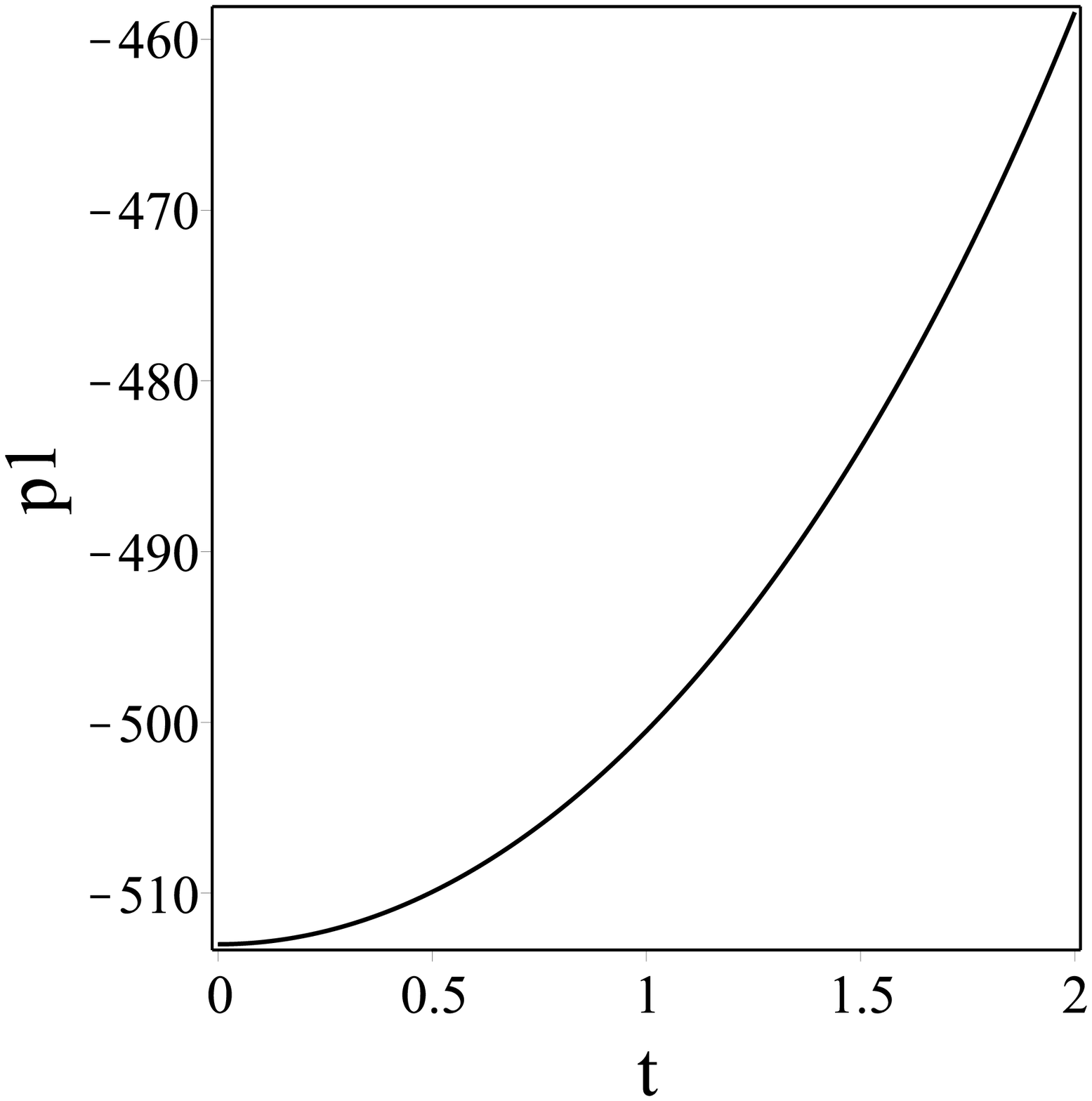}
\caption{Plot of the presure $p_1$ as a function of the cosmic time $t$.} %% ������� � �������
\label{kz8} %% ����� ������� ��� ������ �� ����
\end{minipage}
\hfill
\begin{minipage}[h]{0.4\linewidth}
\includegraphics[width=1\linewidth]{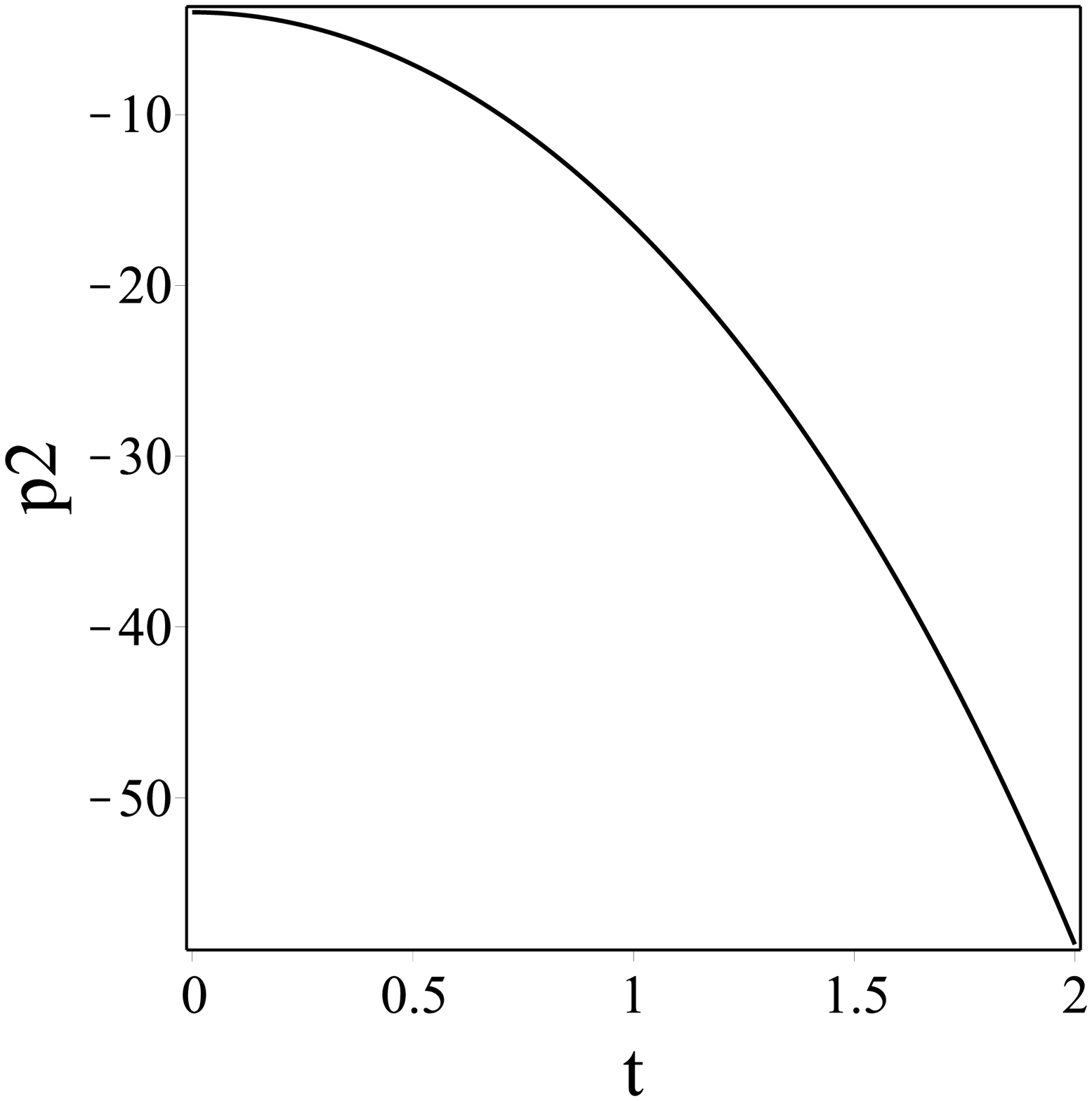}
\caption{Plot of the presure $p_2$ as a function of the cosmic time $t$.}
\label{kz9}
\end{minipage}
\end{center}
\end{figure}

\begin{figure}[h]
\begin{center}
\begin{minipage}[h]{0.4\linewidth}
\includegraphics[width=1\linewidth]{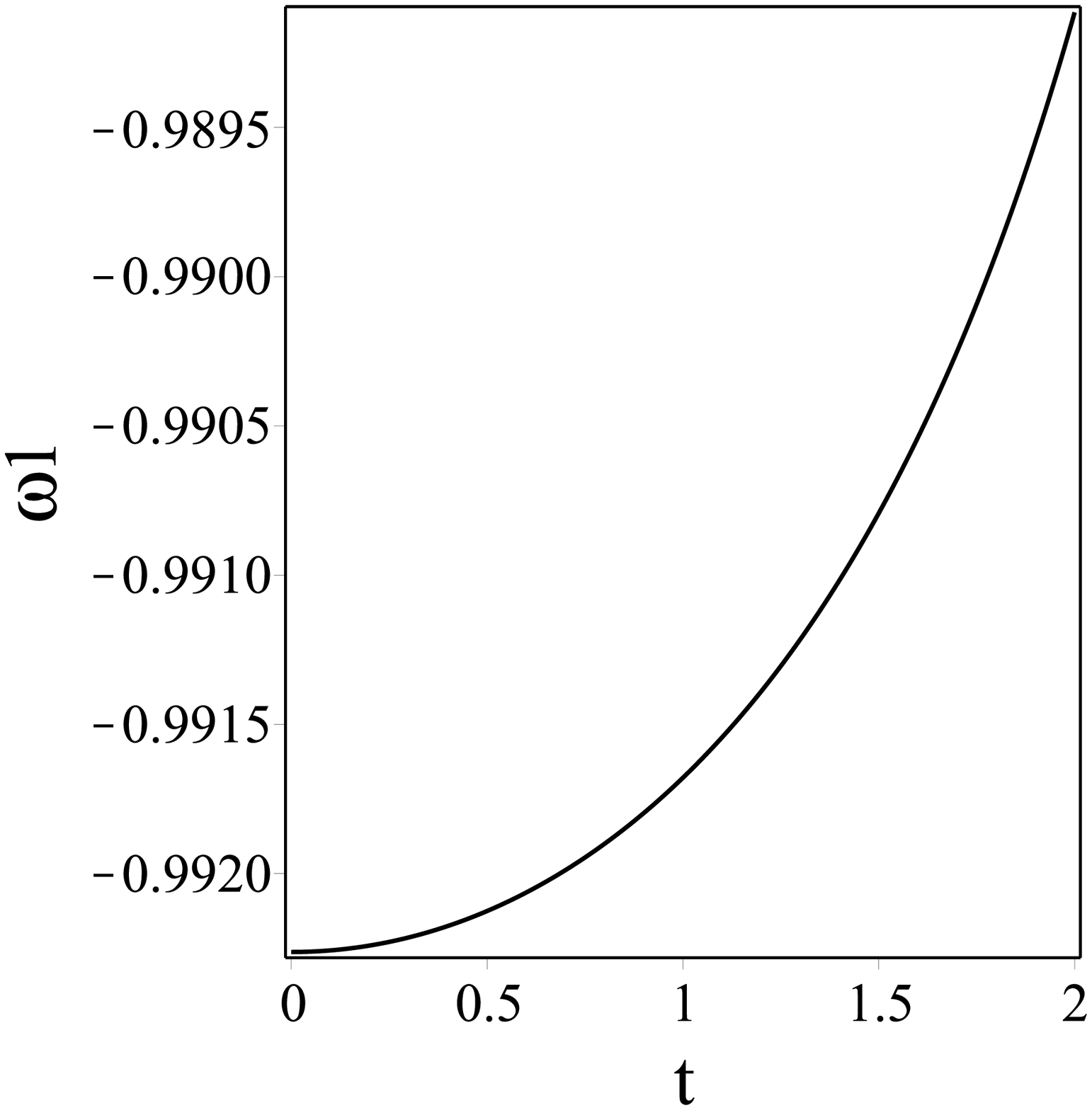}
\caption{Plot of the EoS parameter$\omega_1$ as a function of the cosmic time $t$.} %% ������� � �������
\label{kz10} %% ����� ������� ��� ������ �� ����
\end{minipage}
\hfill
\begin{minipage}[h]{0.4\linewidth}
\includegraphics[width=1\linewidth]{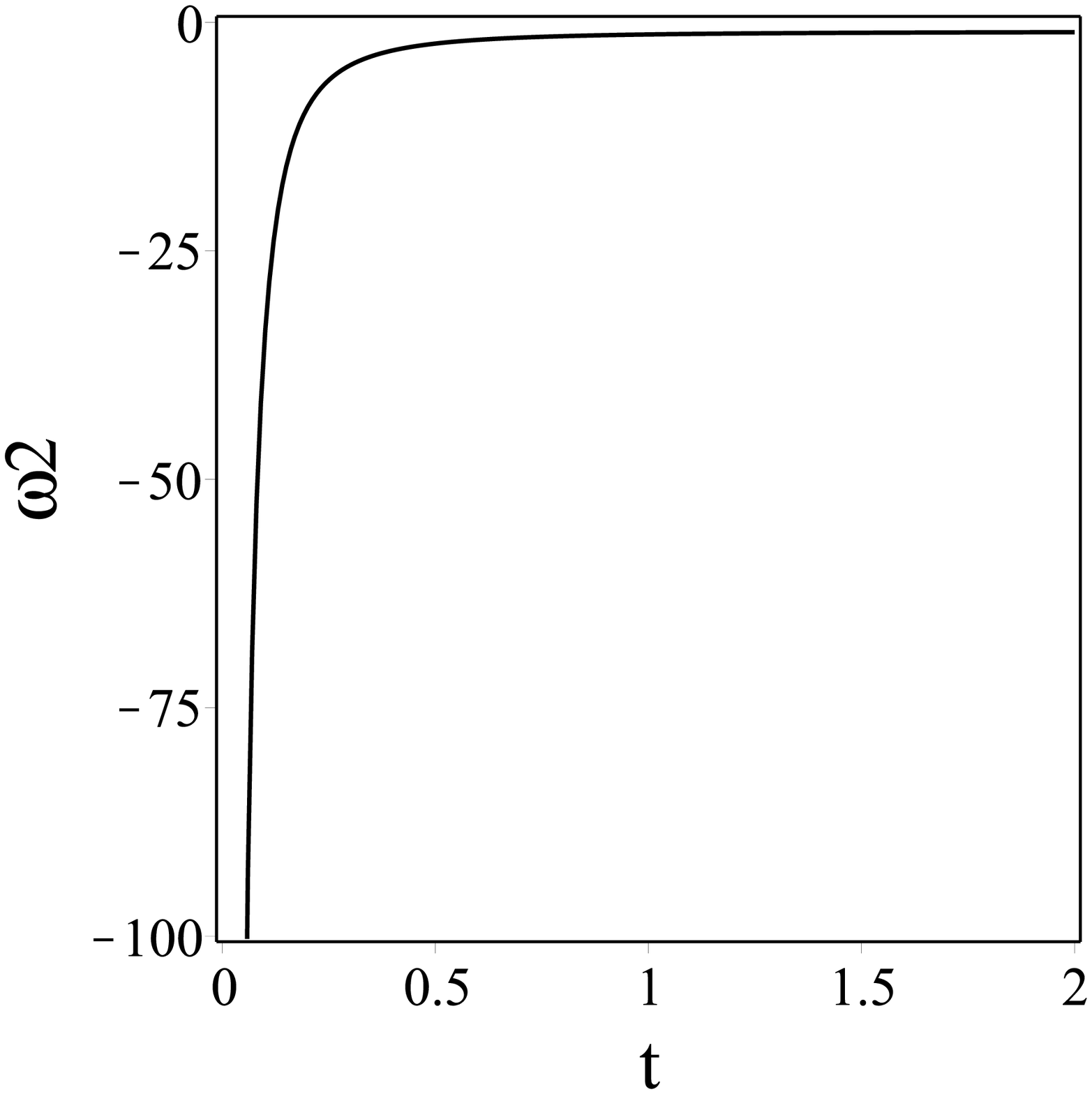}
\caption{Plot of the EoS parameter  $\omega_2$ as a function of the cosmic time $t$.}
\label{kz11}
\end{minipage}
\end{center}
\end{figure}

Moreover, from Eq. (\ref{fe3}), we get the expression for the pressure $p$ and the EoS parameter $\omega$ as follow:
\begin{eqnarray}
p &=&-\left(\rho+\frac{\dot{\rho}}{3H}\right), \label{fe11}\\
\omega &=&\frac{p}{\rho}=-1-\frac{\dot{\rho}}{3H\rho}.\label{fe12}
\end{eqnarray}

Here $H$, $\rho$ and $\dot{\rho}$ are determined by Eqs. (\ref{rat6}), (\ref{fe5}), (\ref{fe6}), (\ref{fe9}) and (\ref{fe10}).
At the end of this Section, we want to show the graphical solution of the equations we have derived. In Figs. \ref{kz6} and \ref{kz7}, we show the cosmological evolution of the energy density $\rho_1$  and energy density $\rho_2$ as a function of $t$. We also depict the cosmological evolutions of the the pressure $p_1$ and the pressure $p_2$ as functions of $t$ in Fig. \ref{kz8} and Fig. \ref{kz9}. Furthermore, in Fig. \ref{kz10} and Fig. \ref{kz11} we demonstrate the cosmological evolution of the EoS parameters $\omega_1$ and $\omega_2$ as a function of the cosmic time $t$. The parameters for the model considered are chosen as $\alpha=1$, $n=2$ and $a_0 = 1$.

%%%%%%%%%%%%%%%%%%%%%%%%%%%%%%%%%%%%%%
\section{The $M_{35}$ - model}
%%%%%%%%%%%%%%%%%%%%%%%%%%%%%%%%%%%%%%%%

In this Section, we want study the $M_{35}$ - model  \cite{VG} which is some  kind  generalizations of the usual FRW LQG.  Its  modified FRW equations are given by:
\begin{eqnarray} \label{fee1}
3H^2&=&8\pi G\rho\sqrt{1-\frac{2\rho}{\rho_c}},\\
\label{fee2} \dot{H}&=&-\frac{4\pi G(\rho+p)}{\sqrt{1-\frac{2\rho}{\rho_c}}}\left(1-\frac{3\rho}{\rho_c}\right),\\
\label{fee3} \dot{\rho}&=&-3H(\rho+p).
\end{eqnarray}
We now can note that if:
\begin{eqnarray} \label{fee1b}
\rho\leq 0.5\rho_{c},\end{eqnarray}
then we get:
\begin{eqnarray} \label{fee1a}
\sqrt{1-\frac{2\rho}{\rho_c}}\approx 1-\frac{\rho}{\rho_c},
\end{eqnarray}
so that Eq. (\ref{fee1}) takes the form of Eq. (\ref{fe1}):
\begin{eqnarray} \label{fee1c}
3H^2=8\pi G\rho\sqrt{1-\frac{2\rho}{\rho_c}}\approx 8\pi G\rho \left(1-\frac{\rho}{\rho_c}\right)
\end{eqnarray}
Now the equation for $\dot{H}$ follows from Eqs. (\ref{fee1}) and (\ref{fee3}). So, the system made by Eqs. (\ref{fee1})-(\ref{fee3}) in the limit given by Eq. (\ref{fee1b}) turns to the usual equations of the standard FRW LQG given in Eqs. (\ref{fe1})-(\ref{fe3}). This is why we tell that the $M_{35}$-model can be considered as some generalization of the usual FRW LQG.

From Eq. (\ref{fee1}), we can obtain following expressions:
\begin{eqnarray}
9H^4 &=&64\pi^2 G^2 \rho^2 \left(1-\frac{2\rho}{\rho_c}\right),\label{fee4}\\
2k\rho^3 &-& k\rho_c \rho^2+9\rho_c H^4=0.\label{fee5}
\end{eqnarray}
Eq. (\ref{fee5}) can be equivalently written as follow:
\begin{equation}
\rho^3-\frac{\rho_c}{2} \rho^2+\frac{9 \rho_c}{2k} H^4=0,\label{fee6}
\end{equation}
where $k=64\pi^2 G^2$. Let us define the new variable $\rho$ as follow:
\begin{eqnarray}
\rho=y+\frac{\rho_c}{6}, \label{fee7}
\end{eqnarray}
Then, Eq. (\ref{fee1a}) can be rewritten as follow:
\begin{eqnarray}
y^3+k_1y+k_2=0, \label{fee7-1}
\end{eqnarray}
where:
\begin{eqnarray}
k_1&=&-\frac{\rho^2_c}{12},\label{fee8}\\
k_2&=&-\frac{\rho^3_c}{108}+\frac{9 \rho_c}{2k}H^4.\label{fee9}
\end{eqnarray}
Then, the three solutions of Eq. (\ref{fee7-1}) are given by:
\begin{eqnarray}
y_1&=&A+B,\label{fee10}\\
y_{2,3}&=&-\frac{A+B}{2}\pm i\frac{A-B}{2}\sqrt{3}, \label{fee11}
\end{eqnarray}
where:
\begin{eqnarray}
A&=&\sqrt[3]{-\frac{k_2}{2}+\sqrt{Q}}, \label{fee12}\\
B&=&\sqrt[3]{-\frac{k_2}{2}-\sqrt{Q}},\label{fee13}\\
Q&=&\left(\frac{k_1}{3}\right)^3+\left(\frac{k_2}{2}\right)^2,\label{fee14}
\end{eqnarray}
which can be also written, using the expressions given in Eqs. (\ref{fee8}) and (\ref{fee9}), as follow:
\begin{eqnarray}
A&=&\sqrt[3]{\frac{\rho^3_c}{216}-\frac{9\rho_c}{4k}H^4+\sqrt{Q}}, \label{fee15}\\
B&=&\sqrt[3]{\frac{\rho^3_c}{216}-\frac{9\rho_c}{4k}H^4-\sqrt{Q}}, \label{fee16}\\
Q&=&-\left(\frac{\rho^2_c}{36}\right)^3+\left(-\frac{\rho^3_c}{216}+\frac{9\rho_c}{4k}H^4\right)^2. \label{fee17}
\end{eqnarray}
We must also remember here that the $i$ given in Eq. (\ref{fee11}) represents the imaginary unit, i.e. $i=\sqrt{-1}$.\\
In our paper, we will consider real solutions only. In the following subsections, we will study two particular cases of the model considered, in particular the case corresponding to $Q=0$ (which corresponds to the de Sitter solution) and the case corresponding to $Q>0$.

\subsection{Case $Q=0$. The de Sitter solution.}
We start considering the first case considered in this work, i.e. the case corresponding to $Q=0$, which yields the de Sitter solution.
In this case, from Eq. (\ref{fee17}), we have:
\begin{eqnarray}
\left( \frac{9 \rho_c}{4k}H^4_0-\frac{\rho^3_c}{216} \right)^2=\left(\frac{\rho^2_c}{36}\right)^3, \label{anto1}
\end{eqnarray}
which yields:
\begin{eqnarray}
\frac{9\rho_c}{4k}H^4_0=\frac{\rho^3_c}{108}, \label{anto2}
\end{eqnarray}
Eq. (\ref{anto2}) can be also rewritten as follow:
\begin{eqnarray}
H^4_0=\frac{k\rho^2_c}{243}. \label{anto3}
\end{eqnarray}
Then, the two solutions of Eq. (\ref{anto3}) will be given by:
\begin{eqnarray}
H_{0+}&=&\left(\frac{k \rho^2_c}{243}\right)^{1/4}, \label{anto4}\\
H_{0-}&=&-\left(\frac{k \rho^2_c}{243}\right)^{1/4}.  \label{anto5}
\end{eqnarray}
In this case, for the scale factor $a$, we have the following expression:
\begin{eqnarray}
a=a_0 e^{H_0 t}.\label{anto6}
\end{eqnarray}
This is a partial case of Eq. (\ref{rat5}). Moreover, for this case, we have that $H=H_0=const$ and $A=B$. Then:
\begin{eqnarray}
y_1 &=& 2A \Rightarrow \ \rho_1=2A + \frac{\rho_c}{6}, \label{anto7}\\
y_2=y_3 &=& -A \Rightarrow \ \rho_2=\rho_3=-A + \frac{\rho_c}{6},\label{anto8}
\end{eqnarray}
where:
\begin{eqnarray}
A=B=\sqrt[3]{\frac{\rho^3_c}{216}-\frac{9\rho^3_c}{4\cdot243}}=-\frac{\rho_c}{3\sqrt[3]{4}}.\label{anto9}
\end{eqnarray}
Then for this case we have following solutions for Eq. (\ref{fee6}):
\begin{eqnarray}
\rho_1 &=&-\frac{2\rho_c}{3\sqrt[3]{4}}+\frac{\rho_c}{6}=-\rho_c\left(\frac{2\sqrt[3]{2}+1}{6}\right), \label{anto10}\\
\rho_2 &=& -\frac{\rho_c}{3\sqrt[3]{4}}+\frac{\rho_c}{6}=\rho_c\left(\frac{2+\sqrt[3]{4}}{6\sqrt[3]{4}}\right).\label{anto11}
\end{eqnarray}
In this case, we also have that:
\begin{eqnarray}
\dot{H}=0.\label{anto12}
\end{eqnarray}

From Eq. (\ref{anto12}) and using Eq. (\ref{fee2}), we can also obtain that:
\begin{eqnarray}
1-\frac{3\rho}{\rho_c}=0, \label{anto13}
\end{eqnarray}
which implies that:
\begin{eqnarray}
\rho=\frac{\rho_c}{3}=const.\label{anto14}
\end{eqnarray}
Then:
\begin{eqnarray}
\dot{\rho}=0.\label{anto15}
\end{eqnarray}
Moreover, from Eq. (\ref{fee3}), we can conclude that:
\begin{eqnarray}
-3H(\rho+p)&=&0, \label{anto16}\\
p&=&-\rho=const, \label{anto17} \\
p&=&-\frac{\rho_c}{3}=const.\label{anto18}
\end{eqnarray}
From Eqs. (\ref{fee1}) and (\ref{fee2}), it follows that:
\begin{eqnarray}
1-\frac{2\rho}{\rho_c}>0,\label{anto19}
\end{eqnarray}
which is equivalent to:
\begin{eqnarray}
\rho<\frac{\rho_c}{2}.\label{anto20}
\end{eqnarray}
So, we have that the energy density $\rho $lies in the interval:
\begin{eqnarray}
0<\rho<\frac{\rho_c}{2}.\label{anto21}
\end{eqnarray}

\begin{figure}[h]
\begin{center}
\begin{minipage}[h]{0.4\linewidth}
\includegraphics[width=1\linewidth]{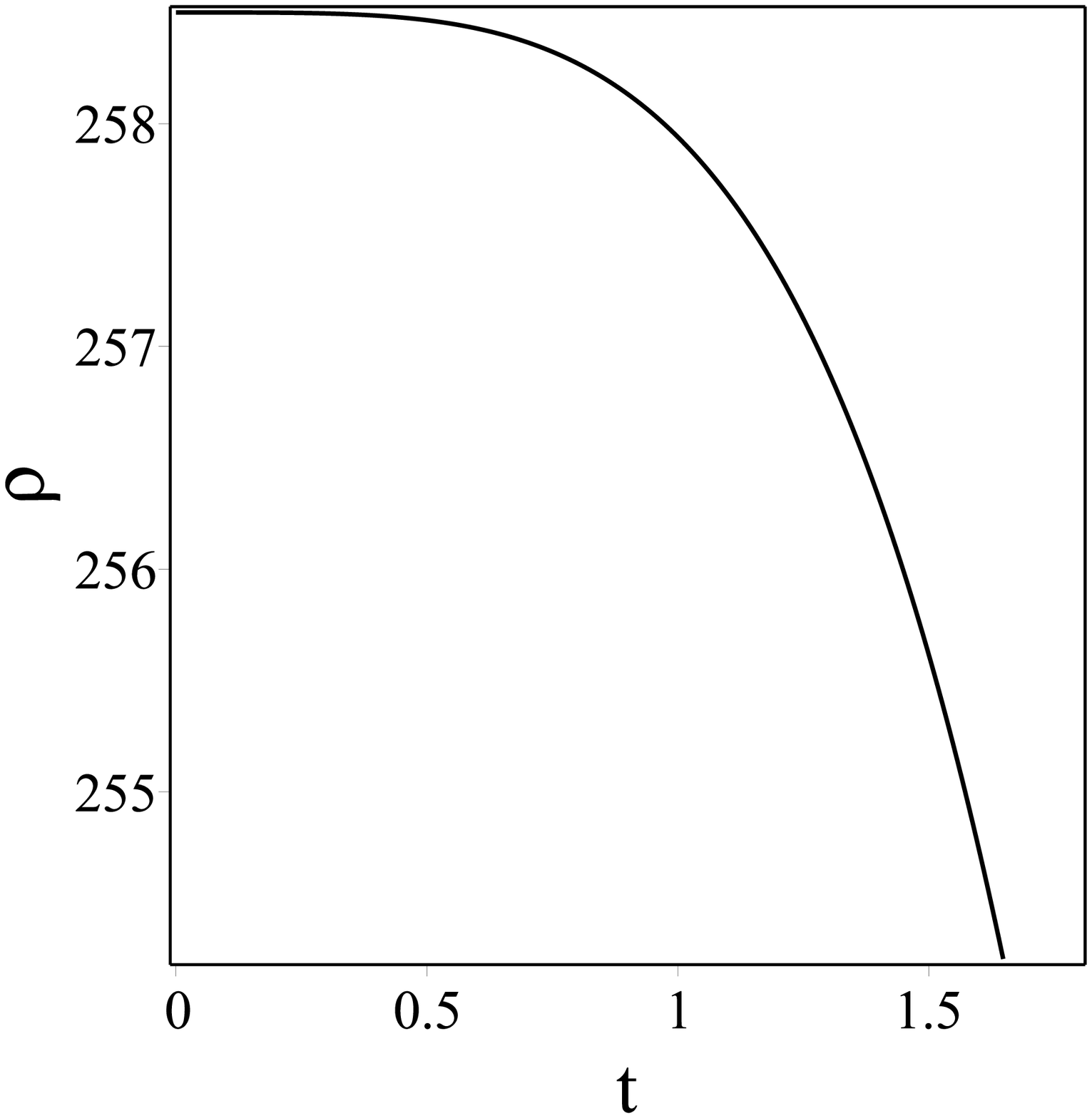}
\caption{Plot of the energy density $\rho$ given in Eq. (\ref{rhoab}) as a function of the cosmic time $t$.} %% ������� � �������
\label{kz12} %% ����� ������� ��� ������ �� ����
\end{minipage}
\hfill
\begin{minipage}[h]{0.4\linewidth}
\includegraphics[width=1\linewidth]{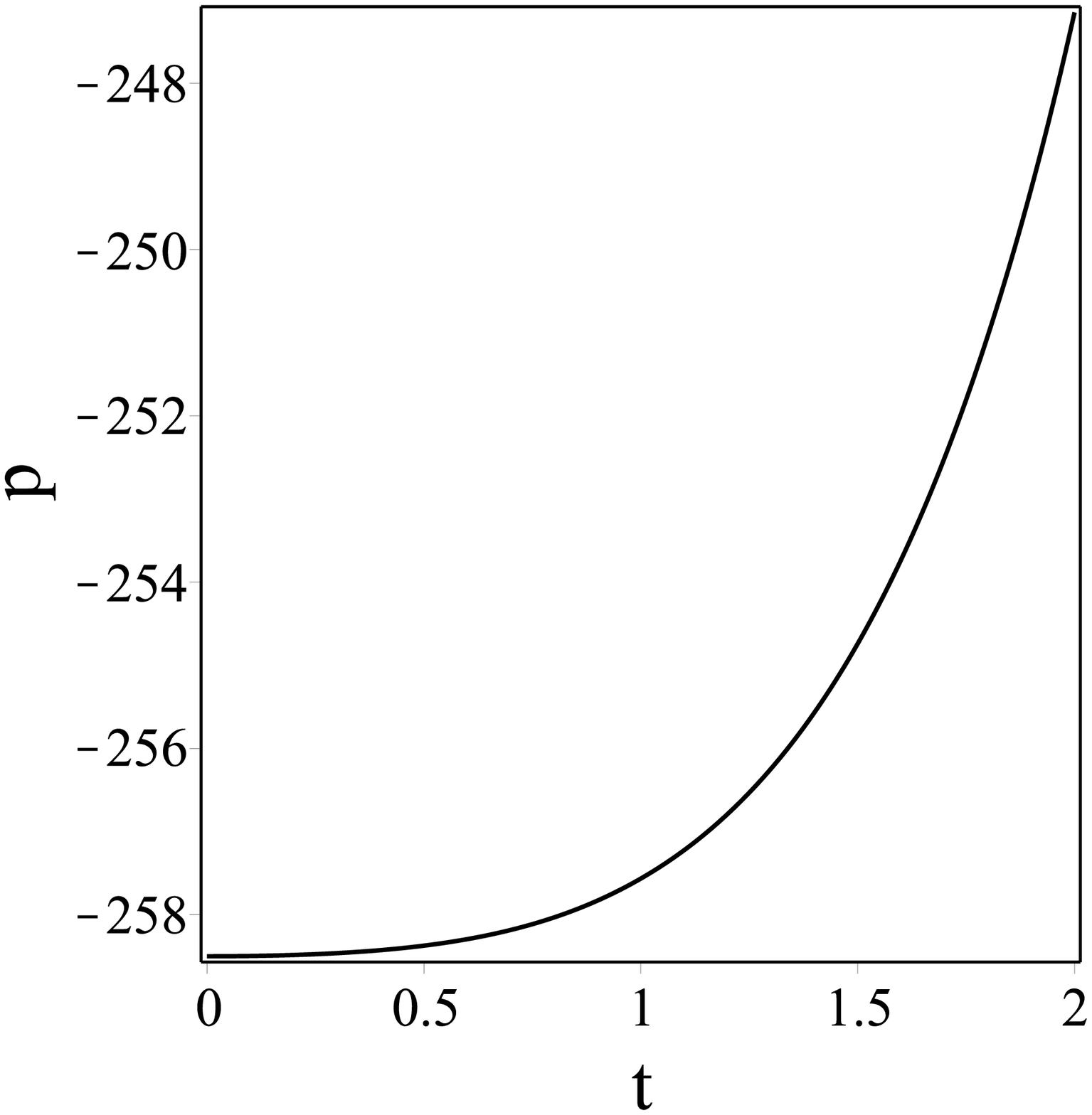}
\caption{Plot of the pressure $p$ for the $M_{35}$ model and $Q>0$ as a function of the cosmic time $t$.}
\label{kz13}
\end{minipage}
\end{center}
\end{figure}

\begin{figure}[h!]
	\centering
		\includegraphics[width=7cm]{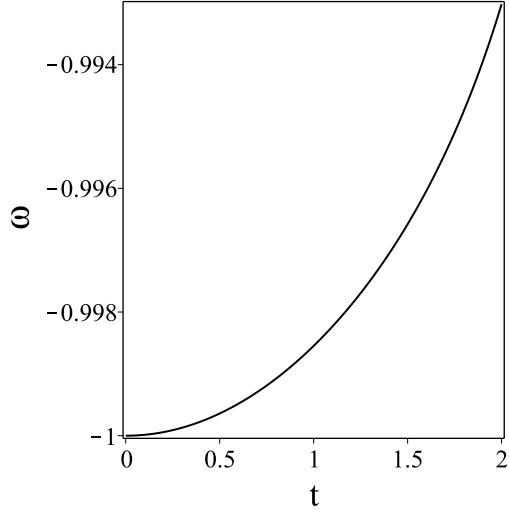}
		\caption{Plot of the EoS parameter $\omega$ for the $M_{35}$ model and $Q>0$ as a function of the cosmic time $t$.} \label{kz14}
\end{figure}

\subsection{Case $Q > 0$}
We now study the second case considered in this paper, which corresponds to $Q>0$.
We decided to consider this case since in Eqs. (\ref{fee15}) and (\ref{fee16}) we should have $Q\geq0$.
For this case, we have get the following conditions for the Hubble parameter $H$:
\begin{eqnarray}
H>H_{0+}
\end{eqnarray}
and
\begin{eqnarray}
H<-H_{0-}.
\end{eqnarray}
Moreover, for this case, we have following solutions:
\begin{eqnarray}
\rho=A+B+\frac{\rho_c}{6}, \label{rhoab}
\end{eqnarray}
where $A$, $B$ and $Q$ are determined in Eqs. (\ref{fee15})-(\ref{fee17}) and $H=H(t)$ function is determined by Eq. (\ref{rat6}).

In Fig. \ref{kz12}, we show the cosmological evolution of the energy density $\rho$ as a function of $t$. We also depict the cosmological evolutions of the the pressure $p$ as functions of the cosmic time $t$ in Fig. \ref{kz13}. Furthermore, in Fig. \ref{kz14}, we demonstrate the cosmological evolution of the EoS parameter $\omega$ as a function of cosmic time $t$. As values of the parameters of the model, we have here chosen $\alpha=1$, $n=2$ and $a_0 = 1$.

%%%%%%%%%%%%%%%%%%%%%%%%%%%%%%%%%%%%%%
\section{Exact cosmological  solutions of the $M_{35}$ - model}
%%%%%%%%%%%%%%%%%%%%%%%%%%%%%%%%%%%%%%%%
In this Section, we will focus our attention to some exact cosmological solutions of the  $M_{35}$ - model we are dealing with, in particular we will study the cosmological properties of the power law-type solution and of the exponential-type solution. Moreover, we will also study a more complicated cases respect to the previous two and we will see which are the results if we consider an arbitrary general function of the time indicated with $f\left( t \right)$.

%%%%%%%%%%%%%%%%%%%%%%%%%%%%%%%%%
\subsection{Power law-type solution}
%%%%%%%%%%%%%%%%%%%%%%%%%%%%%%%%%%

We start considering a particular exact solution of the $M_{35}$ - model, i.e. the power-law one. In this case,  the energy density $\rho$ is given by the following power-law expression:
\begin{eqnarray}
\rho=0.5\rho_{c}(1-\alpha^{2}t^{2n}). \label{5.1}
\end{eqnarray}
Then,  from Eq. (\ref{fee1}), we obtain the following expression for the Hubble parameter $H$:
\begin{eqnarray}
H=\sqrt{\frac{8\pi G\alpha}{3}t^{n}(0.5\rho_{0}-0.5\rho_{0}\alpha^{2}t^{2n})}, \label{5.2}
\end{eqnarray}
We can now find the expression of the pressure $p$ from Eq. (\ref{fee3}), which is also an equivalent from of Eq. (\ref{fe11}), obtaining:
\begin{eqnarray}
p =-\left(\rho+\frac{\dot{\rho}}{3H}\right)=-(0.5\rho_{0}-0.5\rho_{0}\alpha^{2}t^{2n})+\frac{\rho_{0}n\alpha^{2}t^{\frac{3n-2}{2}}}{\sqrt{24\pi G\alpha(0.5\rho_{0}-0.5\rho_{0}\alpha^{2}t^{2n})}}. \label{5.3}
\end{eqnarray}
Finally, for the EoS parameter $\omega$, we derive the following expression:
\begin{eqnarray}
\omega = \frac{p}{\rho}=-1-\frac{\dot{\rho}}{3H\rho}=-1+\frac{\rho_{0}n\alpha^{2}}{\sqrt{24\pi G\alpha}}\frac{t^{\frac{3n-2}{2}}}{(0.5\rho_{0}-0.5\rho_{0}\alpha^{2}t^{2n})^{1.5}}. \label{5.4}
\end{eqnarray}

%%%%%%%%%%%%%%%%%%%%%%%%%%%%%%%%%
\subsection{Exponential solution}
%%%%%%%%%%%%%%%%%%%%%%%%%%%%%%%%%%
We now assume that the energy density $\rho$ has the following exponential form:
\begin{eqnarray}
\rho=0.5\rho_{c}(1-\alpha^{2}e^{2\beta t^{n}}), \label{5.5}
\end{eqnarray}
where $\alpha$, $\beta$ and  $n$ are real contants. In this case, we have:
\begin{eqnarray}
\dot{\rho}=-\alpha^{2}\beta n\rho_{c} t^{n-1}e^{2\beta t^{n}}. \label{5.6a}
\end{eqnarray}
Then, from Eq.(\ref{fee1}), we can easily obtain the following expression for the Hubble parameter $H$:
\begin{eqnarray}
H=\sqrt{\frac{4\pi G\alpha\rho_{c}}{3}e^{\beta t^{n}}(1-\alpha^{2}e^{2\beta t^{n}})}. \label{5.6}
\end{eqnarray}
As done for the power-las case, we can now find the expression of the pressure $p$ from Eq. (\ref{fee3}), obtaining:
\begin{eqnarray}
p =-\left(\rho+\frac{\dot{\rho}}{3H}\right)=-0.5\rho_{c}(1-\alpha^{2}e^{2\beta t^{n}})+\frac{\rho_{0}n\alpha^{2}\beta t^{n-1}e^{1.5\beta t^{n}}}{\sqrt{12\pi G\alpha\rho_{c}(1-\alpha^{2}e^{2\beta t^{n}})}}. \label{5.7}
\end{eqnarray}
Finally for the EoS parameter $\omega$, we obtain the following expression:
\begin{eqnarray}
\omega = \frac{p}{\rho}=-1-\frac{\dot{\rho}}{3H\rho}=-1+\frac{\rho_{c}n\alpha^{2}\beta}{\sqrt{24\pi G\alpha}}\frac{t^{n-1}e^{\beta t^{n}}}{(0.5\rho_{0}-0.5\rho_{0}\alpha^{2}e^{2\beta t^{n}})^{1.5}}. \label{5.8}
\end{eqnarray}

%%%%%%%%%%%%%%%%%%%%%%%%%%%%%%%%%
\subsection{A more complicated solution}
%%%%%%%%%%%%%%%%%%%%%%%%%%%%%%%%%%

We now consider that the energy density $\rho$ can be described by the following expression:
\begin{eqnarray}
\rho=0.5\rho_{c}(1-\alpha^{2}t^{2l}e^{2\beta t^{n}}), \label{5.9}
\end{eqnarray}
where $\alpha$, $\beta$, $l$ and $n$ are real contants. Then, from Eq. (\ref{fee1}), we obtain the following expression for the Hubble parameter $H$:
\begin{eqnarray}
H=\sqrt{\frac{4\pi G\alpha\beta\rho_{c}}{3}t^{l}e^{\beta t^{n}}(1-\alpha^{2}t^{2l}e^{2\beta t^{n}})}. \label{5.10}
\end{eqnarray}
We can now derive the expression of the pressure $p$ for this case using, as before, Eq. (\ref{fee1}), which gives us the following expression for $\dot{\rho}$ for the case considered here:
\begin{eqnarray}
\dot{\rho}=-\rho_{c}\alpha^{2}e^{2\beta t^{n}}(lt^{2l-1}+n\beta t^{2n+l-1}). \label{5.11}
\end{eqnarray}
Furthermore, we get then the following expression for the pressure $p$:
\begin{eqnarray}
p =-0.5\rho_{c}(1-\alpha^{2}t^{2l}e^{2\beta t^{n}})+\frac{\rho_{c}\alpha^{2}e^{2\beta t^{n}}(lt^{2l-1}+n\beta t^{2n+l-1})}{\sqrt{12\pi G\alpha\rho_{c}t^{l}e^{t^{n}}(1-\alpha^{2}t^{2l}e^{2t^{n}})}}. \label{5.12}
\end{eqnarray}
Finally, for the EoS parameter $\omega$, we can easily obtain the following expression:
\begin{eqnarray}
\omega = -1+\frac{\alpha^{2}e^{2\beta t^{n}}(lt^{2l-1}+n\beta t^{2n+l-1})}{(1-\alpha^{2}t^{2l}e^{2\beta t^{n}})\sqrt{3\pi G\alpha\rho_{c}t^{l}e^{\beta\alpha t^{n}}(1-\alpha^{2}t^{2l}e^{2\beta t^{n}})}} \label{5.13}
\end{eqnarray}
or
\begin{eqnarray}
\omega =  -1+\frac{\alpha^{1.5}e^{1.5\beta t^{n}}(lt^{1.5l-1}+n\beta t^{2n+0.5l-1})}{(1-\alpha^{2}t^{2l}e^{2\beta t^{n}})^{1.5}\sqrt{3\pi G\rho_{c}}}.\label{5.14}
\end{eqnarray}

%%%%%%%%%%%%%%%%%%%%%%%%%%%%%%%%%
\subsection{Solution with the arbitrary function $f\left(t\right)$}
%%%%%%%%%%%%%%%%%%%%%%%%%%%%%%%%%%

We now assume that the energy density $\rho$ has the following form:
\begin{eqnarray}
\rho=0.5\rho_{c}\left(1-f^{2}\right), \label{5.15}
\end{eqnarray}
where $f=f\left(t\right)$ is an arbitrary real function of the time $t$ and $f\in [-1, +1]$. Then, from Eq. (\ref{fee1}), we obtain the following expression for the Hubble parameter $H$:
\begin{eqnarray}
H=\sqrt{\frac{4\pi \rho_{c}G}{3}f(1-f^{2})}. \label{5.16}
\end{eqnarray}
In this case, we have the following expression for $\dot{\rho}$:
\begin{eqnarray}
\dot{\rho}=-\rho_{c}f\dot{f}. \label{5.17}
\end{eqnarray}
We can now find the expression of the pressure $p$ from Eq. (\ref{fee3}), getting:
\begin{eqnarray}
p =-\left(\rho+\frac{\dot{\rho}}{3H}\right)=-0.5\rho_{c}(1-f^{2})+\frac{\rho_{c}f\dot{f}}{\sqrt{12\pi \rho_{c}Gf(1-f^{2})}}. \label{5.18}
\end{eqnarray}
Finally, for the EoS parameter $\omega$, we derive the following expression:
\begin{eqnarray}
\omega = \frac{p}{\rho}=-1-\frac{\dot{\rho}}{3H\rho}=-1+\frac{1}{\sqrt{3\pi \rho_{c}G}}(1-f^{2})^{-1.5}\sqrt{f}\dot{f}. \label{5.19}
\end{eqnarray}

%%%%%%%%%%%%%%%%%%%%%%%%%%%%%%%%%%%%%%%%%%%%%%%%%%%%%%%
\section{Scalar field analog of the M$_{35}$-model}
%%%%%%%%%%%%%%%%%%%%%%%%%%%%%%%%%%%%%%%%%%%%%%%%%%%

It is well-known that some cosmological models can be described in the language of scalar field interpretation. We do that here for the M$_{35}$-model. To this aim, let us introduce a scalar field $\phi$ and a self-interacting
potential $U(\phi)$ with the following Lagrangian $ L_{\phi}$:
\begin{eqnarray}
 L_{\phi} = \frac{\dot{\phi}^2}{2} - U(\phi). \label{6.1}
\end{eqnarray}
The corresponding energy-momentum tensor  is equivalent  to a some fluid with energy density $\rho_{\phi}$ and pressure
$p_{\phi}$ so that the Friedmann equations take the following form:
\begin{eqnarray}
3H^2-\rho_{\phi}&=&0,
\label{6.2} \\
 2\dot{H}+3H^{2}+p_{\phi}&=&0,
\label{6.3} \\
\dot{\rho_{\phi}}+3H(\rho_{\phi}+p_{\phi})&=&0,\label{6.4}
\end{eqnarray}
where we put $8\pi G=1$ and:
\begin{eqnarray}
\rho_{\phi} & = & \frac{\dot{\phi}^2}{2} + U(\phi), \label{6.5} \\
p_{\phi} & = & \frac{\dot{\phi}^2}{2} - U(\phi).\label{6.6}
\end{eqnarray}
On the other hand, from Eqs. (\ref{fee1})-(\ref{fee3}), it follows that:
\begin{eqnarray}
3H^2&=&\rho\sqrt{1-\frac{2\rho}{\rho_c}}, \label{6.7}\\
2\dot{H}+3H^{2}&=&-\frac{p(1-\frac{3\rho}{\rho_c})-\frac{\rho^{2}}{\rho_{c}}}{\sqrt{1-\frac{2\rho}{\rho_c}}}, \label{6.8}
\end{eqnarray}
So that, we have:
\begin{eqnarray}
\rho_{\phi}&=&\rho\sqrt{1-\frac{2\rho}{\rho_c}}, \label{6.9}\\
p_{\phi}&=&\frac{p(1-\frac{3\rho}{\rho_c})-\frac{\rho^{2}}{\rho_{c}}}{\sqrt{1-\frac{2\rho}{\rho_c}}}, \label{6.10}
\end{eqnarray}
Hence, we get:
\begin{eqnarray}
\dot{\phi}^2 & = & \frac{(\rho+p)(1-\frac{3\rho}{\rho_c})}{\sqrt{1-\frac{2\rho}{\rho_c}}}, \label{6.11}\\
U(\phi) & = & \frac{\rho(1-\frac{\rho}{\rho_c})-p(1-\frac{3\rho}{\rho_c})}{\sqrt{1-\frac{2\rho}{\rho_c}}}. \label{6.12}
\end{eqnarray}
The Lagrangian $L_{\phi}$ is given by the following relation:
\begin{eqnarray}
 L_{\phi} =  \frac{3p(1-\frac{3\rho}{\rho_c})-\rho(1+\frac{\rho}{\rho_c})}{2\sqrt{1-\frac{2\rho}{\rho_c}}}, \label{6.13}
\end{eqnarray}
so that the FRW-action of the $M_{35}$-model can be written as:
\begin{eqnarray} \label{eh}
S=\int \sqrt{-g}d^4x(R+L_{\phi}),\label{6.14}
\end{eqnarray}
or equivalently as:
\begin{eqnarray} \label{eh}
S=\int \sqrt{-g}d^4x\left[R+\frac{3p(1-\frac{3\rho}{\rho_c})-\rho(1+\frac{\rho}{\rho_c})}{2\sqrt{1-\frac{2\rho}{\rho_c}}}\right].\label{6.15}
\end{eqnarray}

\section{Conclusion}
In this paper, we have considered the classical Friedmann equations for homogeneous and isotropic Friedmann-Robertson-Walker (FRW) models of the Universe
and  generalized the Friedmann equations for Loop Quantum Cosmology (LQC). For these models, the expressions for the energy density have been derived.
For the $M_{35}$-model, the solutions have been obtained for two different cases corresponding to $Q=0$ (which gives the de Sitter solution) and for $Q>0$. The solutions obtained for the models considered (which have been plotted in Figures inserted in the previous Sections)
show that, for the range of values of the parameters considered,  the model studied can describe the accelerated expansion of the Universe.
We have also derived some important cosmological parameters, like the Hubble parameter, the pressure $p$ and the EoS parameter $\omega$, for two exact cosmological solutions of the $M_{35}$- model, in particular the power-law and the exponential solutions. We have also considered a more complicated solution and an arbitrary function of the time $f\left( t\right)$, deriving the same cosmological parameters as for the two exact solutions.
In the last part of the paper, a scalar field description of the model considered is presented  by constructing its self-interacting potential.

\section{Acknowledgement}
S Chattopadhyay acknowledges Visiting Associateship of the Inter-University Centre for Astronomy and Astrophysics (IUCAA), Pune, India.

\end{document}